\documentclass[12pt,a4paper]{article}
\usepackage{amsmath,amssymb,extarrows,mathtools,graphicx,subfigure,setspace}
\usepackage[all]{xy}
\usepackage{epsfig,comment,geometry,comment,color,cite,slashed}
\usepackage[active]{srcltx}
\usepackage[affil-it]{authblk}
\usepackage{fullpage}
\usepackage{subfig}
\usepackage{caption}
\numberwithin{equation}{section}
\newcommand{\be}{\begin{equation}}
\newcommand{\ee}{\end{equation}}
\newcommand{\bea}{\begin{eqnarray}}
\newcommand{\eea}{\end{eqnarray}}

\date{}
\begin{document}
\begin{titlepage}
{\title{\bf\fontsize{14}{15.2}{New Bi-Gravity from New Massive Garvity}}}
\vspace{.5cm}
\author[a]{A. Akhavan}
\author[b]{M. Alishahiha}
\author[a]{A. Naseh}
\author[c]{A. Nemati}
\author[c,a]{A. Shirzad}
\vspace{.5cm}
\affil[a]{ School of Particles and Accelerators, Institute for Research in Fundamental
Sciences (IPM), 
\hspace{.001cm} P.O.Box 19395-5531, Tehran, Iran}
\affil[b]{ School of Physics, Institute for Research in Fundamental
Sciences (IPM), 
\hspace{5.5cm} P.O.Box 19395-5531, Tehran, Iran}
\affil[c]{ Department of Physics, Isfahan University of Technology, 
\hspace{5.5cm} P.O.Box 84156-83111, Isfahan, IRAN}
\renewcommand\Authands{ and }
\maketitle
\vspace{-12cm}
\begin{flushright}
{\small
}
\end{flushright}
\vspace{10cm}
\begin{abstract}
Using the action of three dimensional New Massive Gravity (NMG) we construct a new  
 bi-gravity in three dimensions. This can be done by promoting the rank two auxiliary field appearing 
in the expression of NMG's action into a dynamical field. We show that  small fluctuations around the 
AdS vacuum of the model are non-tachyonic and ghost free within certain range of the 
parameters of the model. We study central charges of the dual field theory and 
observe that in this range they are positive too. This suggests that the proposed model might
be a consistent three dimensional bi-gravity.

\end{abstract}
\end{titlepage}
\setcounter{footnote}{0}
\addtocontents{toc}{\protect\setcounter{tocdepth}{4}}
\setcounter{secnumdepth}{4}
\section{Introduction}

Three dimensional Einstein gravity, with a negative cosmological constant,
 may be thought of as a simplified model to explore and 
understand quantum gravity\cite{Witten:2007kt}. It seems simpler than higher dimensional gravity 
 in the sense that it does not contain local propagating modes, though it is still non-trivial as 
 it has black hole solutions  (BTZ solutions) \cite{Banados:1992wn} and boundary propagating modes. 
A natural question or hope is whether one can understand black hole physics in this model.
Actually three dimensional gravity, and in particular its  BTZ black hole solutions, have 
 increased  our knowledge in understanding the AdS/CFT correspondence
\cite{Maldacena:1997re}. 

To make three dimensional Einstein gravity more realistic, in comparison with higher dimensions,
(in the sense of having local 
propagating modes) one may deform the theory by
higher order derivative terms\footnote{For some recent works on massive gravity theories, see \cite{Cai:2013lqa,Cai:2014upa}.}. In particular one can modify three dimensional Einstein gravity 
by adding a gravitational Chern-Simon term which leads to a  
new theory known as Topologically Massive Gravity (TMG) \cite{Deser:1981wh}. 
Another deformation could be done by adding particular curvature squared terms
to Einstein gravity leading to a new theory;  New Massive
Gravity (NMG)\cite{Bergshoeff:2009hq}. In the latter case one gets two massive 
propagating gravitons\cite{Sadegh:2010pq}; however, in the former one there is only one degree of freedom, due to chirality.

As a result of presence of  higher derivative terms, one expects some 
inconsistencies in the sense that the  model could be tachyonic or could contain ghost. 
Actually for generic values of parameters of the models, the corresponding  equations 
of motion admit several solutions including AdS, BTZ. It is, however, known that these models 
suffer from the fact that the energy of  graviton and the mass of BTZ black holes cannot be 
positive at the same time. More precisely, tuning  parameters to have positive energy for 
graviton results to BTZ solutions with negative mass, and vice versa.
 Equivalently, if one thinks of the 
model as a theory which 
provides a gravitational dual for a two dimensional conformal field theory (CFT), the corresponding 
CFT has negative central charge whenever  the propagating spin-2 mode in the bulk has positive 
energy. This problem is known as  ``bulk-boundary clash''.

It was proposed  \cite{Li:2008dq} that TMG model could become  well defined if 
one tunes the parameters of the model on particular values (critical points), so that the 
modes with negative mass are removed. It is, however, observed 
\cite{Grumiller:2008qz} that at 
the ``critical points''  the model admits new ``logarithmic'' modes which again results to a 
non-unitray theory
 (see also \cite{{Skenderis:2009nt},{Grumiller:2009mw}}). The situation is 
 the same for NMG \cite{{Grumiller:2009sn},{Alishahiha:2010bw}}\footnote{For logarithmic solution 
 in higher dimensional gravity see, e.g.  \cite{{Alishahiha:2011yb},{Gullu:2011sj},
 {Bergshoeff:2011ri}}.}.

In the case of TMG in order to circumvent the above problem, the authors of  ref.
\cite{Bergshoeff:2014pca}  proposed  a new model, named Minimal Massive Gravity (MMG),  
which has the same  
minimal local structure as that of TMG.  Indeed, working within 
the ``Chern-Simons-like'' formulation of massive gravity \cite{Hohm:2012vh} one may find an 
action for the MMG model from that of TMG by adding an extra term. 
Although adding this extra term would open up a possibility to get a consistent ghost free 
and non-tachyonic three dimensional theory, its linearization about a  flat or AdS vacuum still 
has a single massive mode.

The aim of this paper is to explore a possibility of resolving bulk-boundary clash in  the NMG model.
To proceed we note that the action of NMG model 
\bea\label{nmg-1}
I = \frac{1}{16\pi G}\int d^{3}x\sqrt{-g}\hspace{.2mm}\bigg\{
\sigma R-2\Lambda_{g}-\frac{1}{m^{2}}\bigg(R^{\mu\nu}R_{\mu\nu}-\frac{3}{8}
R^{2}\bigg)\bigg\}
\eea
may be recast to the following form, using an auxiliary field $f_{\mu\nu}$ \cite{Bergshoeff:2009hq}
\bea\label{NMG}
I = \frac{1}{16\pi G}\int d^{3}x \sqrt{-g}\left(
\sigma R +f_{\mu\nu}\mathcal{G}^{\mu\nu}+\frac{1}{4}m^2(\tilde{f}^{\mu\nu}f_{\mu\nu}-
\tilde{f}^{2})-2\Lambda_{g}\right),
\eea
where $\sigma = \pm 1$ controls the sign of the curvature term, $G$ is the 
Newton constant, $m$ is a mass parameter,  $ \Lambda_{g} $ is a cosmological parameter,
$R$ is the  Ricci-Scalar and $\mathcal{G}_{\mu\nu}$ is the Einstein tensor of the
metric $g_{\mu\nu}$. Here we have used a notation in which $\tilde{f}^{\mu\nu}
\equiv g^{\mu\alpha}g^{\nu\beta}
f_{\alpha\beta}$, 
$\tilde{f}^{\mu}_{\ \nu}\equiv g^{\mu\alpha}f_{\alpha\nu}$ and
 $\tilde{f}\equiv g^{\mu\nu}f_{\mu\nu}$. Obviously the auxiliary tensor field   $f_{\mu\nu}$ is  not a 
 dynamical field and can be solved using its algebraic equations of motion. Indeed varying the 
 action with respect to $f_{\mu\nu}$ one finds
\bea\label{ff}
f_{\mu\nu} = -\frac{2}{m^{2}} (R_{\mu\nu}-\frac{1}{4}g_{\mu\nu}R).
\eea
Plugging this expression into the action \eqref{NMG} one finds the original  action \eqref{nmg-1}.

The basic idea of our procedure to address the bulk-boundary clash 
 is to promote  the auxiliary field $f_{\mu\nu}$ into a dynamical field by adding a new kinetic 
 term for it to the action. Since the resultant model has  two dynamical spin-2 fields, this theory can be thought of as 
a New Bi-Gravity model (NBG). To be precise we will consider the following action\footnote{Note 
that we have an option to consider the kinetic term of the second metric with either signs
by adding $\tilde{\sigma} R[f]$ with $\tilde{\sigma}=\pm 1$. Of course it what follows we set
$\tilde{\sigma}=1$ and study the model in more details. The results of $\tilde{\sigma}=-1$ 
will be briefly presented in the appendix.\ref{appendixd}.} 
\bea\label{bnmg}
&&I = \frac{1}{16\pi G}\int d^{3}x \sqrt{-g}\left(
\sigma R[g] +f_{\mu\nu}\mathcal{G}^{\mu\nu}[g]+\frac{1}{4}m^2(\tilde{f}^{\mu\nu}f_{\mu\nu}-\tilde{f}^{2})-2\Lambda_{g}\right)\cr\nonumber\\
&&\hspace{3cm}+\frac{1}{16\pi \tilde{G}}
\int d^{3}x\sqrt{-f}\hspace{.5mm}\bigg(R[f]-2\Lambda_{f}\bigg),
\eea
where $ \Lambda_{f} $ is a new 
cosmological parameter, $\tilde {G}$ is the Newton constant of the second metric,
 $R[g]$ and $R[f]$ are Ricci 
scalars constructed from $g_{\mu\nu}$ and $f_{\mu\nu}$, respectively.
${\cal G}_{\mu\nu}[g]$ is also the  Einstein tensor of the metric $g_{\mu\nu}$.  		
It is important to note that all indices  are raised by $g^{\mu\nu}$ except those
in the definition of Ricci scalar $R[f]$ which are raised  by the inverse metric $f^{\mu\nu}$.

It is worth noting that there are other  bi-gravity models in the literature where the authors
have also provided a framework  to address the bulk-boundary clash. In particular motivated 
by the earlier work \cite{Isham:1971gm}  the authors of \cite{Banados:2009it} introduced 
a new three dimensional bi-gravity model. Although one can manage to get a
positive central charge in this model, the model, in fact, is not unitary in the bulk due to the Boulware-Deser ghost.

 Actually our model is very close to another bi-gravity
model introduced in \cite{Bergshoeff:2013xma} to resolve the bulk-boundary clash in NMG.
Although the aim of this paper was to address the same question as ours 
for the NMG model,  an advantage of our model is that it can be written
 using  metric formalism, though the one in \cite{Bergshoeff:2013xma} is formulated  
 using the first order  Dreibein formalism. Of course  the price we paid is that our model 
 contains non-trivial potential given by the Einstein tensor which makes the model more
 complicated to study!

To find the equations of motion of the NBG action one should vary the above action with 
respect to the metrics $g_{\mu\nu}$ and $f_{\mu\nu}$. Doing so one arrives at
\bea \label{e.o.m.g}
&&\sigma\mathcal{G}[g]_{\mu\nu}+\Lambda_{g}g_{\mu\nu}=-\frac{m^{2}}{2}
\bigg[\tilde{f}^{\rho}_{\mu}f_{\nu\rho}-\tilde{f} f_{\mu\nu}-\frac{1}{4}g_{\mu\nu}
(\tilde{f}^{\rho\sigma}f_{\rho\sigma}-\tilde{f}^{2})\bigg]
- 2\tilde{f}_{(\mu}
\hspace{.25mm}^{\rho}\mathcal{G}[g]_{\nu)\rho}\cr\nonumber\\
&&\;\;\;\;\;\;\;\;\;\;\;
-\frac{1}{2}f_{\mu\nu}R[g]+\frac{1}{2}\tilde{f} R_{\mu\nu}[g]+\frac{1}{2}g_{\mu\nu}
f_{\rho\sigma}\mathcal{G}[g]^{\rho\sigma}-\frac{1}{2}\bigg[\nabla^{2}[g]f_{\mu\nu}-
2\nabla[g]^{\rho}\nabla[g]_{(\mu}f_{\nu) 
\rho}\cr\nonumber\\
&&\;\;\;\;\;\;\;\;\;\;\; +\nabla[g]_{\mu}\nabla[g]_{\nu}\tilde{f} +(\nabla[g]^{\rho}\nabla[g]^{\sigma}
f_{\rho\sigma}-\nabla^{2}[g]\tilde{f})g_{\mu\nu}\bigg],\\
&&\mathcal{G}_{\mu\nu}[f]+\Lambda_{f} f_{\mu\nu}=
 \frac{1}{\kappa}
 \sqrt{\frac{g}{f}}\bigg[f_{\alpha\mu}f_{\beta\nu}
\mathcal{G}[g]^{\alpha\beta}+\frac{1}{2}m^{2}\left(g^{\sigma\alpha}
g^{\tau\beta}-g^{\sigma\tau}g^{\alpha\beta}\right)\left
(f_{\sigma\tau}f_{\alpha\mu}f_{\beta\nu}\right)
\bigg],\nonumber
\eea
where ${\cal G}_{\mu\nu}[f]$ is  the  Einstein tensor of the metric $f_{\mu\nu}$ and 
$\kappa= \frac{G}{\tilde{G}}$ is the relative strength of two Newton constants associated with 
two metrics.

Having proposed a model one should first check whether the model contains non-trivial physics.
In order to address this question we will first study different solutions of the above 
equations of motion. Of course for simplicity we will mainly consider solutions where two metrics are 
proportional. Restricted to this set of solutions, we will see that the model admits several 
solutions including AdS vacuums, AdS black hole and AdS wave solutions.  Then 
we will study small fluctuations of the metrics around a vacuum solution, where both metrics are 
AdS whose radius are proportional.

 We will see that within a particular range of the parameters the model is consistent in the sense 
 that, at quadratic level, it is ghost free and non-tachyonic. We will also make a comments on 
 the number of degrees of freedom of NBG. 
Moreover, assuming that the model provides a holographic description for a two
dimensional  CFT, we will compute central charges of the corresponding CFT 
and show that within the same range of parameters they are positive as well. Therefore, our model 
could be a candidate to
address a solution to bulk-boundary-clash in NMG. 

The paper is organized as follows. In the next section we will present different solutions of 
the equations of motion \eqref{e.o.m.g}. In section three we will study the linearized equations 
of motion where we will see that there is a range of  parameters over which the 
model is consistent\footnote{We would like to emphasis that by ``consistent'' we mean 
that the model at the quadratic level is ghost free and non-tachyonic and central charges of 
dual theory are positive.}. In section four we compute central charges for the dual CFT's. The 
last section is devoted to discussions.


\section{Solutions of equations of motion}

In this section we shall investigate some possible solutions of the  equations 
of motion (\ref{e.o.m.g}).  To be concrete, we will first consider 
particular solutions in which two metrics $g_{\mu\nu}$ and $f_{\mu\nu}$ are proportional, {\it i.e.}
$f_{\mu\nu}=\gamma g_{\mu\nu}$. Although this might contains  a small subset of 
 general possible  solutions of the equations of motion,  it is still non-trivial to explore 
different features of the model. We will also present a non-proportional solutions as well.

\subsection{AdS vacuum solution} 

We first investigate the existence of two proportional AdS 
solution 
\bea
ds^{2}_{g} = \frac{\ell_{g}^{2}}{r^{2}} \left(dr^{2}-2dudv\right)
,\hspace{1cm}ds^{2}_{f} = \frac{\ell_{f}^{2}}{r^{2}} \left(dr^{2}-2dudv\right).
\eea
Substituting the above ansatz in the equations of motion   (\ref{e.o.m.g}) one finds 
\bea\label{PRAdS}
\sigma+\ell_{g}^{2}\Lambda_g - \frac{m^2\ell_{f}^{4}}
{4\ell_{g}^{2}}+\frac{\ell_{f}^{2}}{2\ell_{g}^{2}}=0,\;\;\;\;\;\;
&&1-m^{2}\ell_{f}^{2}-\kappa\frac{\ell_{g}}{\ell_{f}}
\left(1+\Lambda_{f}\ell_{f}^{2}\right)=0.
\eea
Assuming $ \ell_{f}^{2}=\gamma \ell_{g}^{2}$, the Eqs.  (\ref{PRAdS}) 
give the following equations to determine  $ \gamma $ and $ \ell_{g}^2 $
\bea\label{ads-p}
\ell^{2}_{g}=-\frac{\sigma +\frac{1}{2}\gamma}{
\Lambda_{g}-\frac{1}{4}m^2 \gamma^{2}}=
\frac{-\kappa+\sqrt{\gamma}}
{\gamma(m^{2}\sqrt{\gamma}+\kappa\Lambda_{f})}.
\eea
From the above relations, we find an equation for $\gamma$ as follows
\bea\label{a}
E[\gamma]\equiv a_{5}\gamma^{5}+a_{4}\gamma^{4}+a_{3}\gamma^{3}
+a_{2}\gamma^{2}+a_{1}\gamma - \Lambda_{g}^{2} =0 ,
\eea
where
\bea
&&a_{1} =\Lambda_{g} 
(\frac{\Lambda_{g}}{\kappa^2}
+2\sigma \Lambda _{f}),\;\;\;\;\;a_{2}=\frac{2\Lambda_{g}\sigma m^{2}}{\kappa^2}
-\Lambda_{f}^2+(\frac{1}{2}
m^2+\Lambda_{f}) 
\Lambda_{g},\cr &&\cr
&&a_{3}= \frac{m^{2}}{\kappa^2}
(m^2+\frac{1}{2}\Lambda_{g})-(\frac{1}
{2}m^2+\Lambda_{f})\sigma\Lambda_{f},\;\;\;\;\;\;a_{4} = \frac{\sigma m^4}{2\kappa^2}
-\frac{1}{4}(\frac{1}{2}m^2+\Lambda_{f})^{2},\cr &&\cr
&&a_{5}= \frac{m^4}{16\kappa^2}.
\eea
Note that the coefficient of $\gamma^5$ in Eq. (\ref{a}) is 
positive, thus the function $E[\gamma]$  approaches  $\pm\infty$ as one  
takes the  limit of $\gamma\rightarrow \pm\infty$.
Moreover one has $E[0]=- \Lambda_{g}^{2}<0$.
Therefore, $E[\gamma]$ would
certainly intersect with the $\gamma$-axis in (at least)
one positive point. This analysis shows that the equations of motion do 
have  proportional AdS solutions.
\subsection{ BTZ black hole solution}

Actually it is clear that since the NBG model admits AdS vacuum solutions, it should also
has BTZ black holes. This is simply because these black holes are locally AdS. Nonetheless in this 
subsection we  will redo the same computations as we did  in the previous subsection for the  
 non-rotating BTZ black hole. Let us consider the following ansatz
\bea
ds_g^{2}=\ell_{g}^{2}\left[(r^2-r_+^2)d\tau^{2}+
\frac{1}{(r^{2}-r_{+}^{2})}dr^{2}+r^{2}d\phi^{2}\right],\;\;\;\;\;\;\;\;ds_f^2=\gamma ds_g^2.
\eea
Plugging this ansatz into the equations of motion  (\ref{e.o.m.g}) one arrives at 
\bea\label{PBTZ}
4 \sigma-\gamma^ 2 \ell_{g}^{2} m^{2}+2 \gamma+ 4 \ell_{g}^{2}  \Lambda_{g}=0,
\;\;\;\;\;\;\;\;\;\;\;
1-m^{2}\ell_{f}^{2}-\kappa\frac{\ell_{g}}{\ell_{f}}
\left(1+\Lambda_{f}\ell_{f}^{2}\right)=0,
\eea
which is indeed the same as those  in the Eq.\eqref{ads-p} as it must be. Actually this is  
the case  simply because 
the solution is locally AdS. Having found the same equations as that in the previous 
subsection one could then argue that the solution does indeed exist.
%
\subsection{Ads wave solutions}
So far we have considered solutions in which two metrics are proportional. 
As an example in this subsection we shall consider a non-proportional solution of the 
equations of motion. To start, we will consider the following AdS wave ansatz for $g_{\mu\nu}$ and $f_{\mu\nu}$
\bea\label{ppwave}
&&ds^{2}_{g} = \frac{\ell_{g}^{2}}{r^{2}} \left(dr^{2}-
2dudv+G(u,r)du^{2}\right),\cr\nonumber\\
&&
ds^{2}_{f} = \frac{\ell_{f}^{2}}{r^{2}} \left(dr^{2}-
2dudv+F(u,r)du^{2}\right).
\eea
Note that this ansatz contains  two free parameters $\ell_g$ and $\ell_f$ and two arbitrary functions
$G(u,r)$ and $F(u,r)$. Now the aim is to find these parameters and functions using the equations 
of motion. Indeed plugging the ansatz into the equations of motion of metric $g_{\mu\nu}$ one finds
an algebraic equation among the parameters of the ansatz and also a partial differential
equation for the functions $G$ and $F$. The corresponding equations are
\bea
\label{PR1}
\ell_{g}^{2}\Lambda_{g}-\frac{1}{4}m^{2}\frac{\ell_{f}^{4}}
{\ell_{g}^{2}}+\sigma+\frac{1}{2}\frac{\ell_{f}^{2}}{\ell_{g}^{2}}=0.
\eea
and 
\bea
\label{de1}
-\left(\frac{\sigma}{2}+\frac{3}{4}\frac{\ell_{f}^
{2}}{\ell_{g}^{2}}\right)\left[\frac{\partial^{2}G}
{\partial r^{2}}-\frac{1}{r}\frac{\partial G}
{\partial r}\right]+\left(\frac{1}{2}\frac
{\ell_{f}^{2}}{\ell_{g}^{2}}\right)\left[\frac
{\partial^{2} F}{\partial r^{2}}
-\frac{1}{r}\frac{\partial  F}{\partial r}
\right]-\left(-\frac{1}{2}\frac{\ell_{f}^{4}}
{\ell_{g}^{2}}m^{2}+\frac{\ell_{f}^{2}}{\ell_{g}^{2}}
\right)\left[\frac{G}{r^{2}}-\frac{ F}{r^{2}}\right]=0.\nonumber\\
\eea
Similarly from the equations of motion of the metric $f_{\mu\nu}$ one gets
\bea
1-m^{2}\ell_{f}^{2}={\kappa}
\frac{\ell_{g}}{\ell_{f}}\hspace{.1mm}(1+\Lambda_{f}\ell_{f}^{2}),
\eea
and
\bea\label{de3}
\frac{\partial^{2}
G}{\partial r^{2}}-\frac{1}{r}\frac{\partial G}
{\partial r}-\kappa \frac{\ell_g}{\ell_f} \left[\frac{\partial^{2} F}{\partial r^{2}}
-\frac{1}{r}\frac{\partial F}{\partial r}\right]
+(2-m^{2}l^{2}_{f})\left
[\frac{G}{r^{2}}-\frac{ F}{r^{2}}\right]=0.
\eea
The differential equations (\ref{de1}) and 
(\ref{de3}) can be solved by considering solutions of the form of $ r^{\alpha} $ 
for functions $G$ and $F$. In particular 
assuming $ \ell_{f}^{2}=\gamma \ell_{g}^{2}$, we 
find a forth order algebraic  equation for $ \alpha $ whose  solutions are 
$\alpha=0,\ 2,\ 1\pm \sqrt{1-A}$ where $A$ is given by 
\bea
A=\frac{\sqrt{\gamma } \left(-\gamma  \ell_{g}^{2} m^2+2\right) 
	\left(2 \sqrt{\gamma } \kappa-
	\gamma +2 \sigma \right)}{ \kappa (3 \gamma +2 
	\sigma )-2  \gamma ^{3/2}}.
\eea
For $\alpha=0$ and $\alpha=2$ we have $F=G$ while for 
$\alpha=1\pm \sqrt{1-A}$ one has $F=\beta G$, where
\bea
\beta=\dfrac{A(\frac{3}{4}\gamma+\frac{\sigma}{2})-\gamma (1-\frac{1}{2}\gamma m^{2} 
\ell_{g}^{2})}{A(\frac{1}{2}\gamma)-\gamma (1-\frac{1}{2}\gamma m^{2} \ell_{g}^{2})}.
\eea 
Putting everything  together one can write the most general solutions for functions $F$ and $G$ as
follows
\bea\label{G1}
&&F(u,r)=f_{1}(u)+f_{2}(u) r^{2}+\beta f_{3}(u) r^{1+\sqrt{1-A}}
+\beta f_{4}(u) r^{1-\sqrt{1-A}},\cr \nonumber\\
&&G(u,r)=f_{1}(u)+f_{2}(u) r^{2}+f_{3}(u) r^{1+\sqrt{1-A}}
+f_{4}(u) r^{1-\sqrt{1-A}},
\eea
where $ f_{i}(u) $'s  are arbitrary functions of $u$.

It is evident from the above solutions that the equations of motion degenerate at certain points where 
the model exhibits logarithmic solutions. More precisely these points are given at  $ A=0$ 
and $A=1$. In particular for $A=0$, from which\footnote{Note that setting $A=0$ one gets 
another solution given by $m^2\ell_g^2\gamma=2$ which, as we will see, should be excluded
due to scalar ghost free condition.} $\sigma-\frac{\gamma}{2}+\kappa\sqrt{\gamma}=0$, 
one finds
\bea\label{Logs}
&&ds^{2}_{g} = \frac{\ell_{g}^{2}}{r^{2}} \bigg(dr^{2}-
2dudv+\left[\tilde{G}_{0}[u]\log(r)
+G_{0}[u]+\tilde{G}_{2}[u]r^{2}\log(r)
+G_{2}[u]r^{2}\right]du^{2}\bigg),\cr\nonumber\\
&&
ds^{2}_{f} = \frac{\gamma \ell_{g}^{2}}{r^{2}} 
\bigg(dr^{2}-2dudv+\left[\tilde{F}_{0}[u]\log(r)
+F_{0}[u]+\tilde{F}_{2}[u]r^{2}\log(r)
+F	_{2}[u]r^{2}\right]du^{2}\bigg).\nonumber\\
\eea 
It is worth noting that due to the logarithmic behavior of the solution they are not asymptotically 
AdS solutions. Indeed from AdS/CFT correspondence one could associate two sources with 
each metric and therefore the corresponding dual theory would be a logarithmic CFT. Having 
two metrics with logarithmic behavior it would be interesting to explore the physical 
content of the dual logarithmic CFT. 

On the other hand for  $A=1$ one finds
\bea\label{sp1}
&&ds^{2}_{g} = \frac{\ell_{g}^{2}}{r^{2}} \bigg(dr^{2}-2dudv+\left[
G_{0}[u]+\tilde{G}_{1}[u]r\log(r)+G_{1}[u]r
+G_{2}[u]r^{2}\right]du^{2}\bigg),\cr\nonumber\\
&&
ds^{2}_{f} = \frac{\gamma \ell_{g}^{2}}{r^{2}} \bigg(dr^{2}-2dudv+\left[
F_{0}[u]+\tilde{F}_{1}[u]r\log(r)+F_{1}[u]r
+F_{2}[u]r^{2}\right]du^{2}\bigg).\nonumber\\
\eea 
For this solution one has 
\bea\label{p1}
m^{2}=\frac{ \kappa (\gamma -2 \sigma )
+4  \sqrt{\gamma } \sigma }{\gamma ^{3/2}
\ell_{g}^2 \left(2 \sigma +2 
  \sqrt{\gamma } \kappa-\gamma\right)}.
\eea
Note that in this case although the solutions still exhibit logarithmic terms, they are, indeed,
asymptotically AdS solutions.

\section{Linearization}\label{lin}

In this section we would like to study small fluctuations of metrics 
around an AdS vacuum solution.  This study may be used to examine the consistency of the NBG 
model. More precisely one could see whether the propagating modes above this 
vacuum is tachyonic or the corresponding modes are ghost, or under which conditions
the resultant model would be ghost free and non-tachyonic.

To proceed let us consider the following AdS vacuum solution 
\be
ds_g^2={\bar g}_{\mu\nu}dx^\mu dx^\nu=\frac{\ell^2}{r^2}(-dt^2+dx^2+dr^2),\;\;\;\;\;
ds_f^2=\gamma ds_g^2,
\ee
and  parametrize the fluctuations above this vacuum as follows
\bea \label{fluc}
g_{\mu\nu}=\bar{g}_{\mu\nu}+h_{\mu\nu},~~~
f_{\mu\nu}=\gamma(\bar{g}_{\mu\nu}+\rho_{\mu\nu}).
\eea
In order to write the quadratic action with respect to the fluctuations one need to expand various 
terms of the action up to second order in the metric perturbations 
$ h_{\mu\nu} $ and $ \rho_{\mu\nu} $. The  details of calculations are given in appendix
 \ref{AppendixB}. It is clear that the linear part of the
action vanishes. On the other hand using the results presented in the appendix 
\ref{AppendixB},  the quadratic part of the action is found
\bea\label{Quad}
&&S^{(2)}[h_{\mu\nu}, \rho_{\mu\nu}]=\frac{1}{16\pi G}\int 
d^{3}x \sqrt{-\bar{g}}\hspace{1mm}\bigg\{(
	\sigma + \frac{3}{2}\gamma)
h^{\mu\nu}(\mathbb{G}^{\ell} h)_{\mu\nu}+\kappa {\sqrt{\gamma}}
  \rho^{\mu\nu}
(\mathbb{G}^{\ell} \rho)_{\mu\nu}
\nonumber\\
&&\hspace{4cm}-{2\gamma}
 h^{\mu\nu}(\mathbb{G}^{\ell}\rho)_{\mu\nu}
+(h-\rho)\cdot (h-\rho)\bigg\},
\eea
where 
$ \mathbb{G}^{\ell} $ is the Pauli-Fierz operator
on the curved $ AdS_{3} $ background with radius 
$\ell$. For  two arbitrary second rank tensors $p$ and $q$ the Pauli-Fierz operator
is
\bea\label{PF}
&&p^{\mu\nu}(\mathbb{G}^\ell q)_{\mu\nu}\equiv -\frac{1}{4}
p_{\nu\rho ;\mu} q^{\nu\rho ;\mu}+\dfrac{1}{2} 
 p_{\mu\nu ;\rho} q^{\rho\nu ;\mu}-\dfrac{1}{4} 
 p_{;\mu} {q^{\mu\nu}}_{;\nu}-\dfrac{1}{4} 
 q_{;\mu} {p^{\mu\nu}}_{;\nu}+\frac{1}{4}p_{;\mu}q^{;\mu}
\nonumber\\
&&\hspace{2.4cm}-\frac{1}{\ell^2}\left
(p^{\mu\nu}q_{\mu\nu}-\frac{1}{2}pq\right),
\eea
where $``\;;\;''$ denotes covariant derivative with respect to the background metric.
The inner product used in Eq.  (\ref{Quad}) is also defined by 
\be
h\cdot \rho\equiv \chi\; h_{\mu\nu}\rho^{\mu\nu}-\xi\; h \rho,
\ee
where $h=\bar{g}^{\mu\nu}h_{\mu\nu}, \rho=\bar{g}^{\mu\nu}\rho_{\mu\nu}, 
\rho^{\mu\nu}=\bar{g}^{\mu\mu'}\bar{g}^{\nu\nu'}\rho_{\mu'\nu'}$ and 
\be\label{MN}
\chi=-\frac{1}{4} m^{2} \gamma^{2} +\frac{1}{2} 
\frac{\gamma}{\ell^{2}},\hspace{1.5cm}\xi=\frac{1}{4}\frac{\gamma}{\ell^{2}}.
\ee
To investigate  whether the model is  ghost free, it is useful to utilize  
a new basis for the fluctuations 
\be\label{basis}
\rho=a\hspace{.5mm}h^{(m)}+ \hspace{.5mm}h^{(0)},\hspace{1cm}
h= h^{(m)}+b\hspace{.5mm}h^{(0)},
\ee
by which  the quadratic part of 
the action  in terms of $h^{(0)}$ and $h^{(m)}$ reads
\bea\label{S2bf}
&&S^{(2)}[h^{(0)},h^{(m)}] = \frac{1}{16\pi G}\int d^{3}x \sqrt{-\bar{g}}\hspace{1mm}
\bigg[\left((\sigma +\frac{3}{2}\gamma)b^{2}+\kappa \sqrt{\gamma}-2\gamma b\right)
h^{(0) \mu\nu}(\mathbb{G}^{\ell}h^{(0)})_{\mu\nu}\cr \nonumber\\
&&\hspace{2cm}+\left(
\sigma +\frac{3}{2}\gamma
+\kappa \sqrt{\gamma}a^{2}-2\gamma a\right)h^{(m) \mu\nu}
(\mathbb{G}^{\ell}h^{(m)})_{\mu\nu}\cr \nonumber \\
&&\hspace{2cm}+{2}\left((\sigma +\frac{3}{2}\gamma)b
+\kappa\sqrt{\gamma}a-\gamma (1+a b)\right)h^{(0) \mu\nu}
(\mathbb{G}^{\ell}h^{(m)})_{\mu\nu}\cr \nonumber\\
&&\hspace{2cm}+(b-1)^{2}\chi \bigg( h^{(0)\mu\nu}
h^{(0)}_{\mu\nu}-\frac{\xi}{\chi} (h^{(0)})^{2}\bigg)\cr \nonumber \\
&&\hspace{2cm}+(a-1)^{2}\chi\bigg( h^{(m)\mu\nu}
h^{(m)}_{\mu\nu}-\frac{\xi}{\chi} (h^{(m)})^{2}\bigg)\cr \nonumber\\
&&\hspace{2cm}-{2}(b-1)(a-1)\chi\bigg(
 h^{(0)\mu\nu}h^{(m)}_{\mu\nu}-\frac{\xi}{\chi} h^{(0)}h^{(m)}\bigg)\bigg].
\eea
The procedure  is  to fix the parameters $a$ and $b$  in such a way that 
the above action reduces to  two decoupled actions for modes $h^{(0)}$ and $h^{(m)}$. 
This can be done
by setting the coefficients of the cross terms (in the third and sixth lines of Eq. \ref{S2bf}) to zero
\be\label{constrain1}
(\sigma +\frac{3}{2}\gamma)b
+\kappa\sqrt{\gamma}a
-\gamma(1+ab) =0,\;\;\;\;\;\;\;\;\;\;\;(b-1)(a-1)=0.
\ee
Moreover to make sure that the resultant action does not have  scalar ghost one should further 
set $\chi=\xi$ \cite{ ArkaniHamed:2002sp,Dvali:2008em} leading to 
$m^2 \ell^2\gamma=1$. Now the aim is to see whether we could find a range for parameters 
of the above action over which it is ghost free and non-tachyonic. 

\begin{figure}
\begin{center}
\centering
\captionsetup{justification=centering,margin=1.5cm}
\includegraphics[height=6cm, width=6.5cm]{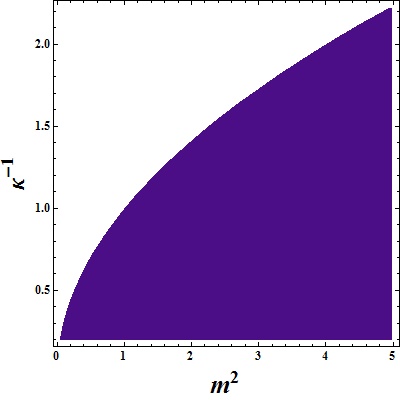}\;\;\;\;\;\;\;\;\;\;\;\;
\includegraphics[height=6cm, width=6.5cm]{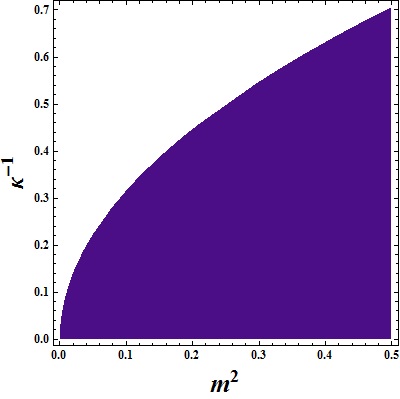}
\caption{Unitary Region for $m^2 $ and $\kappa^{-1}$ for the cases $\sigma=1$ (left) and 
$\sigma=-1$ (right). Note that here we set $\ell=1$.} \label{fig1}
\vspace{0.3cm}
\end{center}
\end{figure}

\begin{figure}
\begin{center}
\centering
\captionsetup{justification=centering,margin=1.3cm}
\includegraphics[height=6cm, width=7cm]{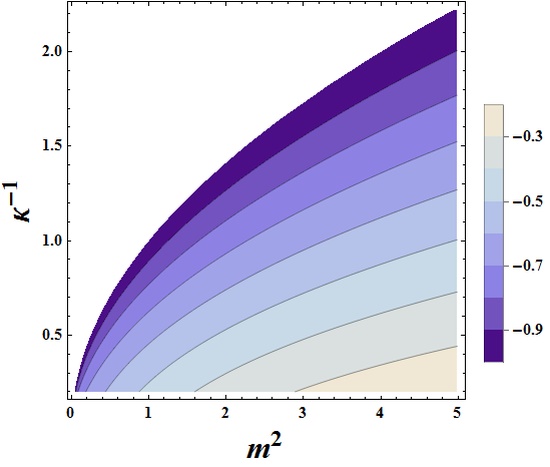}\;\;\;\;\;\;\;\;\;\;\;\;
\includegraphics[height=6cm, width=7cm]{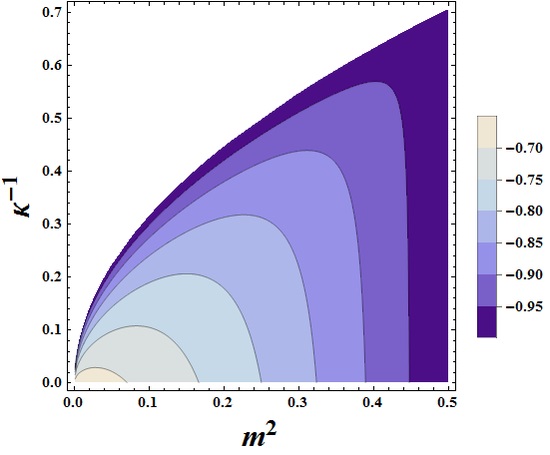}
\caption{Values of $\mathbb{M}^{2}$ within the  unitary region for $m^2 $ and $\kappa^{-1}$
 for the cases
  $\sigma=1$ (left) and $\sigma=-1$ (right). Note that here we set $\ell=1$.} \label{fig2}
\end{center}
\end{figure}

To proceed we note that 
 the solution $a=b=1$ should be discarded. This is because for this solution 
the condition  (\ref{constrain1}) reduces to $\sigma -\gamma/2+\sqrt{\gamma}\kappa = 0$ 
which is exactly the condition  where the NBG model exhibits a logarithmic solution. This fact is also 
clear from the action \eqref{S2bf} as at this point the modes  $h^{(0)}$ and $h^{(m)}$ 
degenerate  resulting to two massless modes. Therefore in what follows we should restrict ourselves
to the cases of $b=1, a\neq 1$ or $a=1, b\neq 1$. 
 It is, however, easy to see that these two cases are 
essentially equivalent and thus it is enough to consider only one of them. In what follows we 
will consider the first case in which $b=1$. In this case   from the equation  (\ref{constrain1}) 
one finds 
\be
a =- \frac{m^{2}\ell^{2}\sigma +\frac{1}{2}}{ \kappa \;m\ell-1}.
\ee
In this case the action (\ref{S2bf}) reads
\bea
&&S^{(2)}[h^{(0)},h^{(m)}] =\frac{1}{16\pi G} \int d^{3}x \sqrt{-\bar{g}}\hspace{1mm}
\bigg[\mathbb{A}_{0}\hspace{.5mm}h^{(0) \mu\nu}
(\mathbb{G}^{\ell}h^{(0)})_{\mu\nu}\hspace{1mm}+\cr \nonumber\\
&&\hspace{2cm}+ \mathbb{A}_{m}\left\{ h^{(m) \mu\nu}
(\mathbb{G}^{\ell}h^{(m)})_{\mu\nu}-\frac{\mathbb{M}^{2}}{4}\bigg( h^{(m)\mu\nu}
h^{(m)}_{\mu\nu}-(h^{(m)})^{2}\bigg)\right\} \bigg],
\eea
where
\bea\label{AA}
&&\mathbb{A}_{0} = 
\sigma -\frac{1}{2m^{2}\ell^{2}}
+\frac{\kappa}{m\ell},\cr\nonumber\\
&&\mathbb{A}_{m} = 
\sigma +(\frac{3}{2}-2a)\frac{1}{m^2 \ell^2}
+\frac{a^2\kappa}{m\ell}
 ,\cr\nonumber\\[2.5mm]
&&\mathbb{M}^{2}=-\frac{1}{\ell^{2}}\;\frac{(a-1)^{2}}{m^2 \ell^2 \mathbb{A}_m}.
\eea
In order to get 
a ghost free and non-tachyonic model certain conditions should be imposed on the parameters 
of  the  action. More precisely 
from  ghost free  condition one  has
\be
\mathbb{A}_0>0,\;\;\;\;\;\;\;\mathbb{A}_m>0,
\ee
while the model is  non-tachyonic if \footnote{It is argued  that for an AdS$_{3}$ vacuum with radius 
$\ell$, the unitarity allows the massive spin-2 to have a negative mass squared $\mathbb{M}^{2}$ 
provided $\mathbb{M}^{2}\geq -1/\ell^{2}$ \cite{Gover:2008sw,Carlip:2008jk,Carlip:2008eq,Bergshoeff:2009aq,Deser:2001us,Deser:2001pe}.} 
\be
\mathbb{M}^2\geq -\frac{1}{\ell^2}.
\ee
One can investigate allowed values for 
$(m^{2}\ell^2,\kappa)$ so that both ghost free and non-tachyonic conditions are satisfied. 
The allowed values of $m^{2}\ell^2$ and $\kappa$ for the 
cases of $\sigma =1$ and $\sigma =-1$ are shown in Fig. \ref{fig1}.  For these values of 
$(m^2\ell^2,\kappa)$ the massive modes acquires  different masses in the range 
$-\frac{1}{\ell^2}\leq \mathbb{M}^2<0$ as depicted in Fig. \ref{fig2}.

For generic value of $\mathbb{M}^2\neq -\frac{1}{\ell^2}$ the spectrum of the NBG 
model consists of a massless graviton mode and a massive spin-2 field which has two 
degrees of freedom. On the other hand when the bound saturates at $\mathbb{M}^2
=-\frac{1}{\ell^2}$ one gets a partially  massive mode which, unlike massive spin-2 field,  
has just  one degree of freedom  at the linear level. Actually  it  has been  shown that the Pauli-Fierz action for this special value of 
 mass has an extra gauge symmetry  which affects the number of degrees of freedom 
 \cite{Deser:2001us,Deser:2001pe,Deser:2001xr, Deser:2004ji,Tekin:2003np}.\footnote{
 Indeed the Pauli-Fierz action for this special value of mass has an gauge symmetry 
 generated by a  scalar gauge parameter $\chi(x)$ as follows \cite{Deser:2001us,Deser:2001pe}
$$
\delta_{\chi}h^{(m)}_{\mu\nu} = \bar{\nabla}_{\mu}\bar{\nabla}_{\nu}\chi -\frac{1}{\ell^{2}}\bar{g}_{\mu
\nu}\chi.
$$
} 
We note, however, that  this gauge symmetry in most  theories is an artifact of linear approximation 
and disappears  at the non-linear level \cite{Blagojevic:2011qc,Alexandrov:2014oda}. 
Of course  if the theory is Weyl invariant, such as 3D conformal gravity, this extra gauge symmetry  
remains in fully non-linear level \cite{Merbis:2014vja}. 
 
In our  case the inequality could saturate for $(a=0, b=1)$ 
in which  the action (\ref{S2bf}) 
reads\footnote{ Note that partial massive gravity could also appear at 
$(a=1,b=0)$ where we get $\kappa=\sqrt{\gamma}$. Although this case 
may be happen for either signs of $\sigma$, the ghost free condition will still be
$\kappa>\sqrt{2}$.}
\bea
&&S^{(2)}[h^{(0)},h^{(m)}] = \frac{1}{16\pi G}\int d^{3}x \sqrt{-\bar{g}}\hspace{1mm}
\bigg[\left(
-2+ \kappa \sqrt{2}
\right)h^{(0) \mu\nu}
(\mathbb{G}^{\ell}h^{(0)})_{\mu\nu}\cr \nonumber\\
&&\hspace{2cm}+2\left\{
h^{(m) \mu\nu}
(\mathbb{G}^{\ell}h^{(m)})_{\mu\nu}+\frac{1}{4\ell^2}\bigg( h^{(m)\mu\nu}
h^{(m)}_{\mu\nu}-(h^{(m)})^{2}\bigg)\right\}\bigg].
\eea
Note that this may happen when $\sigma=-1$ and the ghost free condition reduces to
${\kappa} > \sqrt{2}$.
Therefore at this particular value  the NBG model has a partially massive mode at the linear level. 
We note, however, that since the action (\ref{bnmg}) is not Weyl invariant 
(even for the ghost-free and non-tachyonic parameters), the existence of extra gauge symmetry 
may be an artifact of linear  approximation. It would be interesting to explore
 this point better along the study of 
 \cite{Blagojevic:2011qc, Hassan:2012gz,Hassan:2012rq}.

\section{Central charges of dual theory} 

In this section in order to further investigate the consistency of the NBG 
model we shall study its properties using AdS/CFT correspondence. More precisely the model, 
given by the action \eqref{bnmg}, could provide a holographic description for a dual field theory.
Indeed if one considers the model on an asymptotically locally AdS solution, one would expect 
that the dual theory is a CFT theory, though  since the gravitational theory has two
asymptotically AdS metrics, it is not clear to us what  exactly the corresponding CFT could be! 

Nonetheless, using the free energy of an asymptotically AdS solution
 one may compute the entropy of the corresponding solution  from 
which one can read central charges of the dual theory using Cardy formula for entropy
\cite{Cardy:1986ie,Carlip:2005zn}. Doing so, one
could investigate  conditions on which the obtained central charges are positive. 
Intuitively,  one would expect 
that for a consistent theory the central charges are positive exactly in the range 
of parameters where the model is non-tachyonic and ghost free.

Note that in general on-shell action is divergent and one needs to regularize it by 
adding certain boundary terms to the action. These boundary terms 
are those which are needed to have a well defined variational principle, and those needed 
to remove infinities in the on-shell action. Of course in general finding these 
boundary terms is not an easy task, though in what follows we will find them for a subset 
of solutions where two metrics are proportional.

To proceed we note that  the variation of the action (\ref{bnmg}) 
with respect to $g_{\mu\nu}$ and $f_{\mu\nu}$,
in general, contains  boundary terms which may invalidate
the variational principle for Dirichlet 
boundary condition. These boundary terms emerge from the following 
terms 
\be
\delta I_{\delta\partial_{r}g,\delta\partial_{r}f} = \frac{1}{16\pi G}
\int d^{3}x\left({\sigma} \sqrt{-g} 
\delta R[g]+ \sqrt{-g}f_{\mu\nu}
\delta \mathcal{G}^{\mu\nu}[g]+\kappa \sqrt{-f} \delta R[f]\right),
\ee
which can be written in the following form 
\be\label{Delta}
\delta I_{\delta\partial_{r}g,\delta\partial_{r}f} = \frac{1}{16\pi G}
\int d^{3}x\left(\sqrt{-g}\mathcal
{A}_{\mu\nu}\delta R^{\mu\nu}[g]+
\kappa \sqrt{-f}
f_{\mu\nu}\delta R^{\mu\nu}[f]\right),
\ee
where
\begin{equation}
\mathcal{A}_{\mu\nu}=\sigma g_{\mu\nu}+f_{\mu\nu}-\frac{1}{2}
f_{\alpha\beta}g^{\alpha\beta}g_{\mu\nu}.
\end{equation}
From the first term in Eq. (\ref{Delta}) one gets two different boundary terms as follows
\be\label{Delta.1}
\sqrt{-g}\hspace{1mm}\nabla[g]_{\nu}
(\mathcal{A}^{\mu\nu}\delta \Gamma^{\sigma}_{\mu\sigma}
-\mathcal{A}^{\mu\alpha}\delta \Gamma^{\nu}_{\mu\alpha}),
\ee
and
\bea\label{Delta.2}
&&-\frac{1}{2}\sqrt{-g}\hspace
{1mm}\bigg[\nabla[g]_{\mu}(\nabla[g]_{\nu}\mathcal{A}^{\mu\nu}
g^{\sigma\rho}\delta g_{\sigma\rho})+\nabla[g]_
{\sigma}(\nabla[g]_{\nu}\mathcal{A}^{\mu\nu}g^{\sigma\rho}
\delta g_{\mu\rho})\cr\nonumber \\
&&
-\nabla[g]_{\rho}(\nabla[g]_{\nu}
\mathcal{A}^{\mu\nu}g^{\sigma\rho}\delta g_{\mu\sigma})
\hspace{.3cm}-\nabla[g]_{\mu}(\nabla[g]_{\sigma}
\mathcal{A}^{\mu\nu}g^{\sigma\rho}\delta g_{\nu\rho})
-\nabla[g]_{\nu}(\nabla[g]_{\sigma}\mathcal{A}^{\mu\nu}g^{\sigma
\rho}\delta g_{\mu\rho})\cr\nonumber \\
&&+\nabla[g]_{\rho}(\nabla[g]_{\sigma}\mathcal{A}^{\mu\nu}
g^{\sigma\rho}\delta g_{\mu\nu})\bigg].
\eea
Similarly the same form of boundary terms emerge from the variation of action with 
respect to $f_{\mu\nu}$.

Obviously the boundary terms given in Eq. (\ref{Delta.1}) (also that for $f_{\mu\nu}$)  invalidate the 
variational principle for the Dirichlet 
boundary condition. Of course these terms could be removed by adding  proper 
Gibbons-Hawking terms. Actually since we are interested in solutions where two metrics are 
proportional, $f_{\mu\nu}=\gamma g_{\mu\nu}$, the variation of the action may be recast into 
the following illustrative form
\be
\delta I_{\delta\partial_{r}g,\delta\partial_{r}f} 
= \frac{1}{16\pi G}\int d^{3}x\left(\sqrt{-g}(\sigma-\frac{\gamma}{2}) g_{\mu\nu}\delta 
R^{\mu\nu}[g]+\kappa \sqrt{-f}
 f_{\mu\nu}\delta R^{\mu\nu}[f]\right),
\ee
from which  the  proper Gibbons-Hawking terms can be suggested  as follows 
\be\label{GH}
I_{GH} =- \frac{2(\sigma-\frac{\gamma}{2})}
{16\pi G}\int d^2x\sqrt{-\eta_g}\;
\eta_g^{ij}K[g]_{ij}-\frac{2}
{16\pi \tilde{G}}\int 
d^2x\sqrt{-\eta_f}\;\eta_{f}^{ij}K[f]_{ij},
\ee
where ${\eta_g}_{ij}$ and ${\eta_f}_{ij}$ are the induced metrics, on the boundary, associated 
with the metrics $g$ and $f$, respectively. We would like to re-emphasize that the above boundary 
terms are obtained for  a subclass solutions where two metrics are proportional. Nonetheless it is
worth comparing these terms with the generalized Gibbons-Hawking term which is obtained in  
\cite{Hohm:2010jc} to make the variational principle well defined for the NMG model. 
 Actually one can see that assuming  $f_{ij} = \gamma g_{ij}$, 
the  generalized Gibbons-Hawking terms (given in the Eq.(2.17) of  \cite{Hohm:2010jc}) exactly reduce to the  above boundary term $I_{GH}$ in the limit of  $G \rightarrow 0$.

Let us now evaluate the on-shell action consisting of the original bulk action \eqref{bnmg} 
(named $I_0$) and the above Gibbons-Hawking terms for the following BTZ black hole solution  
\be\label{g}
ds_g^{2}=\ell^{2}\left[(r^2-r_+^2)d\tau^{2}+
\frac{1}{(r^{2}-r_{+}^{2})}dr^{2}+r^{2}d\phi^{2}\right],\;\;\;\;\;\;\;\;ds_f^2=\gamma ds_g^2,
\ee
where $r_+$ is the radius of horizon. Plugging the metrics (\ref{g}) into the action and performing 
integrations over $\tau, r$ and $\phi$ one arrives at 
\be
I_{0} + I_{GH} =\frac{1}{16\pi G} \bigg[\left(a_{\rm div}
+\kappa b_{\rm div}\right){\cal R}^2+(a_{\rm fin}
+\kappa b_{\rm fin})\bigg],
\ee
where ${\cal R}\gg r_+$ is a cutoff and
\bea
&&a_{\rm div} = (2\pi)^2\frac{\ell}{r_+} \left(\sigma
-\frac{3}{4} m^2 \ell^2 \gamma^{2}-\Lambda_{g} 
\ell^2-\frac{1}{2}\gamma\right) ,\cr\nonumber\\
&&b_{\rm div}= (2\pi)^2\frac{\ell}{r_+} \sqrt{\gamma}\left(1
-\Lambda_{f}\gamma \ell^{2}\right),\cr\nonumber\\
&&a_{\rm fin} = (2\pi)^2 \ell r_{+}\left(\sigma +\frac{3}{4}m^2 
\ell^2\gamma^{2}  +\Lambda_{g}  \ell^2
 -\frac{1}{2}\gamma\right), \cr\nonumber\\
&& b_{\rm fin} = (2\pi)^2 \ell r_{+} \sqrt{\gamma}\left(1
+ \Lambda_{f}\gamma \ell^{2} \right).
\eea
It is clear that the on-shell action is divergent due to the infinite volume limit and proper counterterms
are needed to remove the divergent terms. It is easy to see that the corresponding  
boundary counterterms are
\bea
I_{c.t.}= a_{1}\int d^2x\sqrt{\eta _g}+a_{2}\int d^2x\sqrt{\eta_{f}},
\eea
where
\be
 a_{1} =-\frac{1}{16\pi G\ell}\left(\sigma -\frac{3}{4}m^{2}\ell^{2}\gamma^{2}-\Lambda_{g}
 \ell^{2}-
 \frac{\gamma}{2}\right),\;\;\;\;\;\;\; a_{2} = -\frac{1}{16\pi \tilde{G} \ell\sqrt{\gamma}}\left(1-\Lambda_{f}
 \gamma \ell^{2}\right).
\ee
Putting every things together the renormalized on-shell action reads
\begin{equation}
I_{\rm ren} = I_{0}+I_{GH} + I_{ct } = \frac{1}{16\pi G}(
\tilde{a}_{fin}+\kappa \tilde{b}_{fin}),
\end{equation}
with
\bea
&&\tilde{a}_{\rm fin} = 6\pi^2 \ell r_+\left(
\sigma +\frac{1}{4} 
m^2 \ell^2 \gamma^{2}
+\frac{1}{3}\Lambda_{g}  \ell^2-\frac{1}{2}
\gamma\right) ,\cr\nonumber\\
&&\tilde{b}_{\rm fin} =6\pi^2 \ell r_+ \sqrt{\gamma}\left(1
+ \frac{1}{3}\Lambda_{f}\gamma \ell^{2} \right).
\eea
Using the fact that the free energy is given by $F=-T I_{ren}$ with $T=\frac{r_+}{2\pi \ell}$ being the Hawking temperature, the entropy may be found as follows
\bea\label{S.1}
S&=&-\frac{\partial F}{\partial T}\\ &&\cr
&=&
\frac{12\pi^2 \ell r_+}{16\pi G}\left[\bigg( \sigma +\frac{1}{4} m^2 \ell^2\gamma^2 
+\frac{1}{3}\Lambda_{g}  \ell^2-\frac{1}{2}\gamma
\bigg)+ \kappa \sqrt{\gamma}\bigg(1+
\frac{1}{3}\Lambda_{f}\gamma \ell^{2}\bigg)\right].\nonumber
\eea
To read the central charges one should compare this  entropy with  the  Cardy formula 
for entropy given by \cite{Cardy:1986ie}
\be\label{S.2}
S = \frac{\pi^{2} \ell}{3} \left(C_{R} T_{R} + C_{L} T_{L}\right),
\ee
where $C_{L(R)}$ and $T_{L(R)}$ are central charge and temperature of left (right) moving sector
respectively. On the other hand since the model under consideration is parity invariant one has 
\be\label{T}
C_{R} = C_{L} \equiv C,\;\;\;\;\;\;\;\;\;\; 
T_{R} = T_{L} = \frac{r_{+}}{2\pi \ell}.
\ee
Therefore comparing Eq.  (\ref{S.1}) with (\ref{S.2}) and using Eq. (\ref{T}) one arrives at
\be\label{c}
C=\frac{1}{16\pi G}(C_k+\kappa C_{\tilde{k}}),
\ee
where
\be \label{c1}
C_{k} = 36\pi \ell\left(\sigma +\frac{1}{4}m^2 \ell^2 \gamma ^2+\frac{1}{3} 
\Lambda_{g}  \ell^2-\frac{1}{2}\gamma \right),\;\;\;\;\;\;
C_{\tilde{k}} = 36\pi \ell\sqrt{\gamma }\left(1
+\frac{1}{3} \Lambda_{f}\gamma \ell^{2}\right).
\ee
It is important to note that since the metrics given in the Eq. \eqref{g} are solutions of
 the equations of motion, their parameters should be fixed from Eq. (\ref{PBTZ}).
Actually from the equation \eqref{PBTZ} and taking 
into account the scalar ghost free condition \footnote{Note that 
since BTZ solution is locally AdS, the scalar ghost free condition in the present case is the same as
that of AdS solution; {\it i.e.} $\gamma m^2\ell^2=1$.} one finds
\bea\label{lambdag}
\Lambda_{g}=-\frac{1}{\ell^{2}}\Big(\sigma +\dfrac{1}{4m^{2}\ell^{2}} \Big),\;\;\;\;\;\;\;\;
\Lambda_{f}= -m^2,\;\;\;\;\;\;\;\;\;
\gamma=\frac{1}{m^2\ell^2}.
\eea
Plugging these expressions into the equation \eqref{c1} one  arrives at
\bea\label{centralcharge}
C=\frac{ 3  \ell}{2 G}\left( \sigma -\frac{1}{2 m^{2}\ell^{2}}
+\frac{\kappa}{m\ell}\right),
\eea
which is the central charge for the CFT dual of the NBG model.
Obviously for $\kappa^2\rightarrow 0$ where the model reduces to NMG  one  gets
\bea
C = \frac{3\ell}{2G} (\sigma-\frac{1}{2 m^{2}\ell^{2}}),
\eea
which is the well-known central charge of the CFT dual to the NMG 
model on an asymptotically locally 
$AdS_{3}$ geometry \cite{Bergshoeff:2009aq,Liu:2009kc}. Since the way we have obtained 
the above expression for central charge crucially depends on the boundary terms
 and also due to its important role to address  the bulk-boundary clash, 
it would be  interesting if one could find this expression from another consideration. Indeed 
in the Appendix.\ref{AppendixC} we have re-derived this expression using entropy function formalism by which
one may compute the entropy of an extremal black hole \cite{Sen:2005wa}.

Interestingly enough using  the equation \eqref{AA} the above central charge may be 
recast into the following form
\be\label{CA}
C=\frac{3 \ell}{2G}\; \mathbb{A}_0,
\ee
where $\mathbb{A}_0$ is the coefficient of the kinetic term of the massless graviton
which is positive within the range of our interest (see Fig.\ref{fig1}). Therefore the absence of scalar ghost and positivity of kinetic terms, trivially implies that the central charge is positive. Note that 
in this range $\mathbb{A}_m$ is also positive. Thus the NBG model could resolve the bulk-boundary clash of the NMG model. 

It is worth noting  that the central charge may be also recast into the
form of $C=\frac{3\ell}{2G}(\sigma-\frac{\gamma}{2}+\kappa\sqrt{\gamma})$ from which it is 
clear that setting $C=0$ one arrives at a logarithmic CFT (see also Eq. \eqref{Logs}).
 
The  relation between central charge and the coefficient of the kinetic term of 
massless graviton (equation \eqref{CA}) could be simply understood from the fact that  
the  massless graviton is 
the source for the dual stress-energy tensor. Note also that there  is no  a cross term 
between massless graviton and spin-2 field and therefore  two-point functions of 
stress-energy tensor can be entirely obtained  from the action of massless graviton. On the 
other hand, since the corresponding  two-point 
functions are proportional to the dual central charge,  the positivity of 
central charge should be encoded into the  positivity of the kinetic term in the action of massless graviton. 

Therefore if, for a model,  one could provide a condition that both $\mathbb{A}_0$ and 
$\mathbb{A}_m$ are positive, this would automatically give positive central charge and resolve
bulk-boundary clash too. This is, indeed, the case for the NBG model.

\section{Conclusions }

In this paper we have addressed the bulk-boundary clash for NMG model by proposing a new
bi-gravity in three dimensions. To introduce our model we have inspired by NMG action 
written in terms of a rank two auxiliary field. The idea was to promote the auxiliary field to 
a dynamical field.

We have shown that the resultant model given by the action \eqref{bnmg} has non-trivial content 
in the sense that its equations of motion admit several non-trivial solutions such as BTZ black 
hole and AdS waves. Of course we have mainly considered a small subset of all possible 
solutions where two metrics are proportional. In particular we have considered a vacuum solution where both 
metrics are AdS geometers  whose radii are proportional, {\it i.e.} $\ell^{2}_f=\gamma \ell^{2}_g$.

We have also studied small fluctuations around this AdS vacuum where we have shown that
at linear level the model consists of a massless graviton and a massive spin-2 field. We have 
shown that there is a range for parameters where the NBG model becomes ghost 
free non-tachyonic and therefore could be classically consistent. 
It is important to note that for these values of 
parameters the model has the same degrees of freedom as that of NMG. 

We note, however, that 
for a particular value of parameters the spin-2 field becomes partially massive which 
means that in the linearized level it has just one degree of freedom. Of course we would expect 
that this reduction of degrees of freedom is an artifact of the linear approximation. Accepting 
this fact we would like to conclude that the NBG model has the same degrees of freedom 
as that of  NMG and therefore could be thought of as a model which 
addresses solution of bulk-boundary clash for NMG.

Intuitively it is simple to see how our model could resolve bulk-boundary clash. Indeed we 
have shown that upon linearization the quadratic part of the action contains two parts: one 
for a massless graviton and one for a massive spin-2 field. Denoting the coefficients of these 
terms  by $\mathbb{A}_0$ and $\mathbb{A}_m$, respectively, one can see that for NMG model 
these two factors cannot be positive. Moreover since $\mathbb{A}_0$ is related to the central 
charge of the dual theory one generally  gets the  bulk-boundary clash. Whereas for 
the NBG model there is a range of parameters where both $\mathbb{A}_0$
and $\mathbb{A}_m$ can be positive simultaneously. This range of parameters are given 
in Fig.\ref{fig1}. 
 
Of course it is important to note that our study crucially depends on the fact that we have 
considered a vacuum AdS solution where two metrics are proportional. The theory could have 
other solutions and linearization around them would be different from that considered in this paper.
It will be interesting to study these solutions and fluctuations around them. 
It is also worth noting that in this paper we have only studied the consistency of the model up 
to quadratic terms of the action. To fully address the problem one should also consider higher 
order terms to make sure that the theory remains ghost free. Alternatively one may study the 
constraint system of the full theory using Hamintonian formalism. We hope to address this question
in our  future work.

Finally we note that  although the way we have 
found the central charge of the dual theory relies on the particular solution we have 
taken,  we think that  the expression given in the equation  
(\ref{centralcharge}) is general and could be found by another method such as
 asymptotic symmetry analysis of the model.
 
\section*{Acknowledgments}
We would like to thank H. Afshar,  E. Altas, A. F. Astaneh, E. Bergshoeff , S. Carlip, D. Grumiller, A. Mollabashi, M. R. Mohammdai Mozaffar, F. Omidi, S. F. Taghavi and M. R. Tanhayi for useful discussions.    A. N. would also like to thank ICTP where the project is finalized for warm hospitality. We would also like to thank the referee for his/her comments.

\section*{Appendix}

\appendix

\section{Details of Variation of Action}\label{AppendixA}
In this appendix we present details of the variation of the
action (\ref{bnmg}). 
For variation with respect to $g^{\mu\nu}$ of various terms we have
\bea
&&\delta\sqrt{-g}=\frac{-1}{2}\sqrt{-g}
g_{\mu\nu}\delta g^{\mu\nu},\cr\nonumber\\
&&\delta (\sqrt{-g}R[g])=\sqrt{-g}\hspace{1mm}G_{\mu\nu}[g]\delta 
g^{\mu\nu}+\mathcal{B}_{1},\cr\nonumber\\
&&\delta(\tilde{f}^{\mu\nu}G_{\mu\nu})=\delta g^{\mu\nu}
\bigg(2 \tilde{f}_{(\mu}~^{\rho}G_{\nu)\rho}
+\frac{1}{2}\bigg[\nabla^{2}[g]f_{\mu\nu} -2\nabla[g]^{\rho}\nabla[g]_{(\mu}f_{\nu)\rho}
+ \nabla[g]_{\mu}\nabla[g]_{\nu}\tilde{f}\cr\nonumber\\
&&\hspace{3.4cm}+(\nabla[g]^{\rho}\nabla[g]^{\sigma}f_{\rho\sigma}
-\nabla^{2}[g]\tilde{f})g_{\mu\nu}\bigg]+\frac{1}{2}R f_{\mu\nu}-
\frac{1}{2}\tilde{f} R_{\mu\nu}\bigg)
+\mathcal{B}_{2},\cr\nonumber\\
&&\delta(\tilde{f}^{\mu\nu}f_{\mu\nu}-\tilde{f}^{2})=2
(f_{\nu\rho}\tilde{f}_{\mu}^{\rho}-\tilde{f}f_{\mu\nu})
\hspace{.5mm}\delta g^{\mu\nu},
\eea
where the boundary term $\mathcal{B}_{1}$ is
\bea
\mathcal{B}_{1} = \sqrt{-g}\hspace{1mm}\nabla[g]_{\nu}
\bigg(g^{\mu\nu}\delta \Gamma^{\sigma}_{\mu\sigma}[g]
-g^{\mu\alpha}\delta \Gamma^{\nu}_{\mu\alpha}[g]\bigg),
\eea
and the boundary term $\mathcal{B}_{2}$, setting  $B_{\mu\nu}=
f_{\mu\nu}-\frac{1}{2}fg_{\mu\nu}$, is 
\bea
&&\mathcal{B}_{2}=\sqrt{-g}\hspace{1mm}
\nabla[g]_{\nu}\bigg(B^{\mu\nu}\delta \Gamma^{\sigma}_{\mu\sigma}
-B^{\mu\alpha}\delta \Gamma^{\nu}_{\mu\alpha}\bigg)
-\frac{1}{2}\sqrt{-g}\hspace{1mm}
\bigg(\nabla[g]_{\mu}(\nabla[g]_{\nu}
B^{\mu\nu}g^{\sigma\rho}\delta 
g_{\sigma\rho})\cr\nonumber\\
&&\hspace{1cm}+\nabla[g]_{\sigma}(\nabla[g]_
{\nu}B^{\mu\nu}g^{\sigma\rho}\delta 
g_{\mu\rho})-\nabla[g]_{\rho}(\nabla[g]_
{\nu}B^{\mu\nu}g^{\sigma\rho}\delta 
g_{\mu\sigma})-\nabla[g]_{\mu}(\nabla[g]_{\sigma}
B^{\mu\nu}g^{\sigma\rho}\delta 
g_{\nu\rho})\cr\nonumber\\
&&\hspace{1cm}-\nabla[g]_{\nu}(\nabla[g]_{\sigma}
B^{\mu\nu}g^{\sigma\rho}\delta g_{\mu\rho})
+\nabla[g]_{\rho}(\nabla[g]_{\sigma}B^{\mu\nu}
g^{\sigma\rho}\delta g_{\mu\nu})\bigg).
\eea
Variations of different terms in the 
action (\ref{bnmg}) with respect to $ f^{\mu\nu} $ are
\bea
&&\delta\sqrt{-f}=-\dfrac{1}{2}\sqrt{-f}
f_{\mu\nu}\hspace{.5mm}\delta f^{\mu\nu},\cr\nonumber\\
&&\delta (\sqrt{-f}R[f])=\sqrt{-f} 
\hspace{.5mm}G_{\mu\nu}[f]\delta f^{\mu\nu}
+\mathcal{B}_{3},\cr\nonumber\\
&&\delta (f_{\mu\nu}G^{\mu\nu}[g])
=-f_{\mu\alpha}f_{\nu\beta}G^{\mu\nu}[g]\hspace{.5mm}\delta 
f^{\alpha\beta},\cr\nonumber\\
&&\delta (\tilde{f}^{\mu\nu}f_{\mu\nu}-\tilde{f}^{2})=
-2\bigg(g^{\mu\alpha}g^{\nu\beta}-g^{\mu\nu}
g^{\alpha\beta}\bigg)f_{\alpha\beta}f_{\mu\alpha}
f_{\nu\beta}\delta f^{\alpha\beta},
\eea
where the boundary term $\mathcal{B}_{3}$ is
\bea
\mathcal{B}_{3}=\sqrt{-f}\hspace{1mm}
\nabla[f]_{\nu}\bigg(f^{\mu\nu}\delta \Gamma^{\sigma}_{\mu\sigma}[f]
-f^{\mu\alpha}\delta \Gamma^{\nu}_{\mu\alpha}[f]\bigg).
\eea
\section{Details of Linearization}\label{AppendixB}
In this appendix we present the detailed calculations of linearization of
different terms in the action (\ref{bnmg}). To do so we will consider the following 
perturbations around an AdS$_{3}$ solution 
%
\bea\label{down}
g_{\mu\nu}=\bar{g}_{\mu\nu}+h_{\mu\nu},\hspace{1cm}
f_{\mu\nu}=\gamma(\bar{g}_{\mu\nu}+\rho_{\mu\nu}),
\eea
where $ \bar{g}_{\mu\nu} $ is the metric of the 
$ AdS_{3} $ solution with the radius $\ell$. 
Assuming $g^{\mu\nu}$ and $f^{\mu\nu}$ as the inverse tensors corresponding to the metrics $g_{\mu\nu}$ and $f_{\mu\nu}$ respectively
we have the contravariant version of Eq. (\ref{down}) as follows
\bea\label{up}
g^{\mu\nu}=\bar{g}^{\mu\nu}-h^{\mu\nu}+
h^{\mu}_{\lambda}h^{\lambda\nu},\hspace{1cm}f^{\mu\nu}=\gamma^{-1} 
(\bar{g}^{\mu\nu}-\rho^{\mu\nu}+
\rho^{\mu}_{\lambda}\rho^{\lambda\nu}),
\eea
where  $h^{\mu\nu}\equiv \bar{g}^{\mu\alpha}\bar{g}^{\nu\beta}h_{\alpha\beta}$ and  $\rho^{\mu\nu}\equiv \bar{g}^{\mu\alpha}\bar{g}^{\nu\beta}\rho_{\alpha\beta}$.
The zeroth, first and second order of different terms of the action (\ref{bnmg}) with respect to the above perturbations are given by 
\bea
&&(\sqrt{-g}R[g])^{(0)}=\sqrt{-\bar{g}}\hspace{1mm}\left
(\frac{-6}{\ell^{2}}\right),\cr\nonumber\\
&&(\sqrt{-g} R[g])^{(1)}=\sqrt{-\bar{g}}\hspace{1mm}\bigg(-h_{;\mu}
^{~\mu}+h_{\mu\sigma}^{~~;\mu\sigma}-\frac{1}{\ell ^2}h\bigg),\cr\nonumber\\
&&(\sqrt{-g}R[g])^{(2)}=\sqrt{-\bar{g}}\hspace{1mm}
\bigg(-\frac{1}{4}h_{\nu\rho;\mu}h^{\nu\rho;\mu}+
\frac{1}{2}h_{\mu\nu;\rho}h^{\mu\rho;\nu}
-\frac{1}{2}h_{;\mu}h^{\mu\nu}_{~~;\nu}
\cr\nonumber\\
&&\hspace{4.1cm}+\frac{1}{4}h^{;\mu}h_{;\mu}-\frac{1}
{2 \ell^2}(h^{\mu\nu}h_{\mu\nu}-\frac{1}{2}h^2)\bigg).
\eea
\bea
&&(\sqrt{-f}R[f])^{(0)}=
\gamma^{\frac{1}{2}}\sqrt{-\bar{g}}
\left(\frac{-6}{\ell^{2}}\right),\cr\nonumber\\
&&(\sqrt{-f}R[f])^{(1)}=\gamma^{-1}\sqrt{-\bar{g}}\hspace{.5mm}
\left(-\rho_{;\mu}^{~\mu}+\rho_{\mu\sigma}^{~~;\mu\sigma}-\dfrac{1}{\ell^2}
\rho\right),\cr\nonumber\\
&&(\sqrt{-f}R[f])^{(2)}=\gamma^{\frac{1}{2}}\sqrt{-\bar{g}}\hspace{1mm}
\bigg(-\frac{1}{4}\rho_{\nu\alpha;\mu}\rho^{\nu\alpha;\mu}
+\frac{1}{2}\rho_{\mu\nu;\alpha}\rho^{\mu\alpha;\nu}
-\frac{1}{2}\rho_{;\mu}\rho^{\mu\nu}_{~~;\nu}
\cr\nonumber\\
&&\hspace{4.65cm}+\frac{1}{4}\rho^{;\mu}\rho_{;\mu}-
\frac{1}{2 \ell^2}(\rho^{\mu\nu}
\rho_{\mu\nu}-\frac{1}{2}\rho^2)\bigg).
\eea
\bea
&&(\sqrt{-g}f_{\mu\nu}G[g]^{\mu\nu})^{(0)}=
\gamma\sqrt{-\bar{g}}\left(\frac{3}{\ell^{2}}\right),\cr\nonumber\\
&&(\sqrt{-g}f_{\mu\nu}G[g]^{\mu\nu})^{(1)}=
\gamma\sqrt{-\bar{g}}\left(\frac{1}{\ell^{2}}
(\rho-\frac{1}{2}h)-\frac{1}{2} (-h_{;\mu}^{~\mu}
+h_{\mu\sigma}^{~~;\mu\sigma})\right),\cr\nonumber\\
&&(\sqrt{-g}f_{\mu\nu}G[g]^{\mu\nu})^{(2)}=
{\frac{\gamma}{2}}
\sqrt{-\bar{g}}\hspace{.5mm}\bigg(
h^{\nu\sigma}_{~~;\nu} h_{\mu\sigma}^{~~;\mu}-h^{\nu\sigma}_
{~~;\nu}h_{;\sigma}+ 
h_{\nu}^{\sigma} h_{\mu\sigma ;}^{~~~~\mu\nu}
+\frac{1}{4}h^{;\mu}h_{;\mu}\cr\nonumber\\
&&\hspace{3cm}+\frac{1}{2}h_{\mu\sigma;\nu}h^{\mu\nu;\sigma}
-\frac{3}{4}h_{\mu\sigma;\nu}h^{\mu\sigma;\nu}
-h^{\nu\sigma}h_{\rho\sigma;\nu}^{~~~~\rho}-
\frac{1}{2} h h_{;\mu}^{~\mu}\cr\nonumber\\
&&\hspace{3cm}+h(\frac{1}{2}h^{\mu\nu}_{~~;\mu\nu}+
\frac{3}{4\ell^{2}}h)+\rho^{\mu\nu}\left(-h_{\mu\nu;\sigma}^{~~~~\sigma}-
h_{;\mu\nu}+2 h_{\mu\sigma;\nu}^{~~~~\sigma}\right)\cr\nonumber\\
&&\hspace{3cm}+\rho(h_{;\mu}^{~\mu}-h^{\mu\nu}_{~~;\mu\nu}-
\frac{1}{\ell^{2}}h)+\frac{2}{\ell^2}h_{\mu\nu}
\rho^{\mu\nu}-\frac{3}{2\ell^2}h_{\mu\nu}h^{\mu\nu}\bigg).
\eea
\bea
&&(\sqrt{-g} \tilde{f}^{\mu\nu}f_{\mu\nu})^{(0)}=3 
\gamma^{2}\sqrt{-\bar{g}},\:\:\:\:(\sqrt{-g} \tilde{f}^{\mu\nu}f_{\mu\nu})^{(1)}=\gamma^{2}
\sqrt{-\bar{g}}
\left(2\rho-\frac{h}{2}\right),\cr\nonumber\\
&&(\sqrt{-g} \tilde{f}^{\mu\nu}f_{\mu\nu})^{(2)}=\gamma^{2}\sqrt{-\bar{g}}
\left(\frac{9}{4} h^{\mu\nu}h_{\mu\nu}-4  
h^{\mu\nu}\rho_{\mu\nu}+ \rho^{\mu\nu}
\rho_{\mu\nu}-\frac{5}{8}h^{2}+h\rho\right).
\eea
\bea
&&(\sqrt{-g}\tilde{f}^2)^{(0)}=9 \gamma^{2}\sqrt{-\bar{g}},\:\:\:\:\:
(\sqrt{-g}\tilde{f}^2)^{(1)}=\gamma^{2} \sqrt{-\bar{g}} 
\left(6\rho-\dfrac{3}{2}h\right),\cr\nonumber\\
&&(\sqrt{-g}\tilde{f}^2)^{(2)}=\gamma^{2}\sqrt{-\bar{g}}
\left(\rho h+\rho^{2}-\frac{7}{8}h^{2}+\frac{15}{4} 
h^{\mu\nu}h_{\mu\nu}-6h^{\mu\nu}\rho_{\mu\nu}\right).
\eea
\bea
&&(\sqrt{-g}\Lambda_{g})^{(0)}=\sqrt{-\bar{g}}
\Lambda_{g},\;\;\;\;\;\;\;(\sqrt{-g}\Lambda_{g})^{(1)}=\sqrt{-\bar{g}}
\left(\frac{h}{2}\right)\Lambda_{g} ,\cr\nonumber\\
&&(\sqrt{-g} \Lambda_{g})^{(2)}=\frac{1}{8}\sqrt{-\bar{g}}
\left(h^{2}-2 h^{\mu\nu}h_{\mu\nu}\right)\Lambda_{g}.
\eea
\bea
&&(\sqrt{-f}\Lambda_{f})^{(0)}=\gamma^{\frac{3}{2}}
\sqrt{-\bar{g}}\Lambda_{f},\;\;\;\;\;\; (\sqrt{-f}\Lambda_{f})^{(1)}=\gamma^{\frac{3}{2}}
\sqrt{-\bar{g}}\left(\frac{\rho}{2}\right)\Lambda_{f},\cr\nonumber\\
&&(\sqrt{-f} \Lambda_{f})^{(2)}=\frac{1}{8}\gamma^{3/2
}\sqrt{-\bar{g} }\left(\rho^{2}-2\rho^{\mu\nu}
\rho_{\mu\nu}\right) \Lambda_{f}.
\eea

\section{Central charge from entropy Function}\label{AppendixC}

In this section we  would like to provide another way to find the central charge (\ref{centralcharge})
using Sen entropy function procedure\cite{Sen:2005wa}. In fact it was proved that for an extremal 
black hole whose near horizon geometry develops an AdS$_2$ geometry the entropy can be
found by minimizing the following entropy function
\be
S=2\pi(e q-f)
\ee
where $f=-2\pi \int_{\rm horizon}\mathcal{L}$, with $\mathcal{L}$ is Lagrangian density evaluated on the solution.  The entropy is given by the minimal value of entropy function (see  \cite{Sen:2005wa}).
Here $e$ is a parameter of the solution whose conjugate is $q$ (see below for more details).

Let us now consider an extremal BTZ black hole in our model. The corresponding geometry can 
be express as AdS$_2$ fibration of AdS$_3$ whose metric may be parametrized as follows
\bea\label{ansatz}
ds^{2}_{g} = v_{1}\left(-r^2 dt^2+\frac{dr^2}{r^2}\right) +v_{2}
\left(d\phi+ e\hspace{.5mm}r\hspace{.5mm} dt\right)^2,\hspace{1cm}ds^{2}_{f} = \gamma
\hspace{1mm}ds^2_{g}. 
\eea
Of course for this metric to present an extremal BTZ black hole one should have 
$v_2=\frac{v_1}{e^2}$ and $v_1=\frac{\ell^2}{4}$. But for the moment we keep them two free  independent parameters. These values should come
 automatically as one minimizes the entropy function. Note that the above metric is in
the form of $AdS_{2}\times S^{1}$.

Now using the ansatz (\ref{ansatz}) and plugging it into the the Lagrangian density obtained from the 
action  (\ref{bnmg}), one finds
\bea\label{EF}
S = 2\pi e q +\frac{\pi}{G} \left(S_{G} + \kappa S_{\tilde{G}}\right),
\eea 
where
\bea
&& S_{G} = \frac{e^{2}v_{2}^{\frac{3}{2}}}{16v_{1}}\left(\gamma -2\sigma\right)+\left(\left[\frac{3}{8}m^{2}\gamma^{2}+\frac{1}{2}\Lambda_{g}\right]v_{1}-\frac{1}{4}(\gamma -2\sigma)\right)\sqrt{v_{2}},\cr \nonumber\\
&& S_{\tilde{G}} = -\frac{e^{2} v_{2}^{\frac{3}{2}}}{8 v_{1}}\sqrt{\gamma}
+\frac{1}{2}\left(\gamma \Lambda_{f} v_{1} +1\right) \sqrt{\gamma v_{2}}\hspace{.5mm}.
\eea
By extremizing the entropy function (\ref{EF}) with respect to $v_{1},v_{2}$ and $e$, one gets
\bea\label{Ext}
&& E^{v_{1}} \equiv E^{v_{1}}_{G} +\kappa E^{v_{1}}_{\tilde{G}} =0,\cr
&& E^{v_{2}} \equiv  E^{v_{2}}_{G} + \kappa E^{v_{2}}_{\tilde{G}}=0 ,\cr 
&& E^{e} \equiv 2\pi q +\frac{\pi}{G} \left(E^{e}_{G} + \kappa E^{e}_{\tilde{G}}\right)=0,
\eea
where
\bea
&& E^{v_{1}}_{G} = -\frac{e^{2}v_{2}^{\frac{3}{2}}}{16v_{1}^{2}}\left(\gamma -2\sigma\right)+\left(\frac{3}{8}m^{2}\gamma^{2}+\frac{1}{2}\Lambda_{g}\right)\sqrt{v_{2}},\;\;\;\;
 E^{v_{1}}_{\tilde{G}} = \frac{e^{2} v_{2}^{\frac{3}{2}}}{8 v_{1}^{2}}\sqrt{\gamma}
+\frac{1}{2}\gamma^{\frac{3}{2}}\Lambda_{f}\sqrt{v_{2}},\cr \nonumber\\
&&  E^{v_{2}}_{G} = \frac{3 e^{2}\sqrt{v_{2}}}{32v_{1}}\left(\gamma -2\sigma\right)+\left(\left[\frac{3}{16}m^{2}\gamma^{2}+\frac{1}{4}\Lambda_{g}\right]v_{1}-\frac{1}{8}(\gamma -2\sigma)\right)\frac{1}{\sqrt{v_{2}}},\\
&& E^{v_{2}}_{\tilde{G}} =  -\frac{3 e^{2} \sqrt{v_{2}}}{16 v_{1}}\sqrt{\gamma}
+\frac{\sqrt{\gamma}}{4}\left(\gamma \Lambda_{f} v_{1} +1\right) \frac{1}{ \sqrt{v_{2}}},
\;\;\;\; E^{e}_{G} = \frac{e v_{2}^{\frac{3}{2}}}{8 v_{1}} (\gamma -2\sigma),\;\;\;\; E^{e}_{\tilde{G}} = -\frac{e v_{2}^{\frac{3}{2}}}{4 v_{1}}\sqrt{\gamma}.\nonumber
\eea
From the last equation in (\ref{Ext}) we have
\bea\label{e}
q =\frac{(\sqrt{\gamma}\kappa +\sigma-\frac{\gamma}{2})v_{2}^{\frac{3}{2}}}{8G v_{1}}
\;e ,
\eea
while from the two other we arrive at
\bea
E^{v_{1}}-2\frac{v_{2}}{v_{1}}E^{v_{2}} = \frac{\pi \sqrt{v_{2}}}{2Gv_{1}^{2}}
\left( \sqrt{\gamma}\kappa+\sigma -\frac{\gamma}{2}\right)\left(e^{2}v_{2}-v_{1}\right) =0.
\eea
Clearly $ \sqrt{\gamma}\kappa+\sigma -\frac{\gamma}{2}$ can not be zero otherwise we 
get $q=0$ and a log gravity. Therefore the only solution of the above equation is  
\bea\label{v2}
v_{2} = \frac{v_{1}}{e^{2}},
\eea
that is what was anticipated. One can then find a solution for $v_{1}$ by plugging the above 
expression of $v_{2}$ into the  Eq.(\ref{Ext})
\bea\label{v1}
v_{1} = -\frac{\sqrt{\gamma}\kappa +\sigma -\frac{\gamma}{2}}{4\kappa \gamma^{\frac{3}{2}}\Lambda_{f} +4\Lambda_{g}+3m^{2}\gamma^{2}}.
\eea 
On the other hand since the solution is locally AdS one has 
\bea\label{parameters}
\Lambda_{g} =-\frac{1}{\ell ^{2}}\left(\sigma +\frac{1}{4m^{2}\ell ^{2}}\right),
\hspace{1cm}\Lambda_{f} = -m^{2},\hspace{1cm} \gamma = \frac{1}{m^{2}\ell^{2}}.
\eea
Note that using these values for the parameters of the model from \eqref{v1} one gets $v_1=
\frac{\ell^2}{4}$ as expected. Finally putting every things together and using the equations 
 (\ref{e}),(\ref{v1}) and (\ref{v2}) one arrivers at 
\bea
S = \frac{\pi \ell}{4 G e} \left(\sigma + \frac{\kappa}{m \ell} -\frac{1}{2 m^{2}\ell^{2}}\right) =  2\pi \sqrt{\frac{q\ell}{4G}\left(\sigma -\frac{1}{2m^{2} \ell^{2}}+  \frac{\kappa}{m \ell}\right)}.
\eea
which is the entropy of our extremal BTZ black hole that can be recast in to the following  Cardy formula 
\be
S=\frac{\pi^2}{3}T C,
\ee
where the central charge $C$ and Frolov-Thorne temperature $T$ are given by
\be
T = \frac{2}{\pi}\frac{\sqrt{G q}}{\sqrt{\ell\left(\sigma -\frac{1}{2m^{2} \ell^{2}}+  \frac{\kappa}{m \ell}
\right)}},\;\;\;\;\;\;
C =\frac{3\ell}{2G}\left(\sigma -\frac{1}{2m^{2} \ell^{2}}+  \frac{\kappa}{m \ell}\right),
\ee
in agreement with the equation (\ref{centralcharge}). The identification of the 
temperature may also be understood from the period of the compact direction in the ansatz 
\eqref{ansatz} (see for example \cite{Guica:2008mu}).


\section{$\tilde{\sigma}=-1$ case}\label{appendixd}

As we have already mentioned one could have considered either sign for the 
kinetic term of the metric $f_{\mu\nu}$. This option could be taken into account 
by multiplying the corresponding terms by $\tilde{\sigma}=\pm 1$ then the 
the action for metric $f_{\mu\nu}$ reads
\be
\frac{1}{16\pi \tilde{G}}\int d^3x\sqrt{f}\bigg(\tilde{\sigma}R[f]-2\Lambda_f\bigg).
\ee
The main parts of our paper have devoted to the case of  $\tilde{\sigma}=1$. In this appendix we 
will  briefly present the results of  $\tilde{\sigma}=-1$ case. Indeed classical solutions we have 
found  in section two can be maped to this case by replacing $\kappa\rightarrow-\kappa$ and $
\Lambda_f\rightarrow -\Lambda_f$. .

Of course one needs to be more careful when we want to study the linearized  action 
above a vacuum. Actually starting with the fluctuations \eqref{fluc} and going through all
computations and using the same field redefinition as \eqref{basis} one arrives at
\be
b=1,\;\;\;\;\;\;\;\;\;\;a=\frac{m^2\ell^2\,\sigma+\frac{1}{2}}{m\ell\,\kappa+1}.
\ee
Therefore, from the positivity of $\gamma$ we have $\sigma =-1$, in which the quadratic action reads
\bea\label{S2}
&&S^{(2)}[h^{(0)},h^{(m)}] = \frac{1}{16 \pi G}\int d^{3}x \sqrt{-\bar{g}}\hspace{1mm}
\bigg[\mathbb{A}_0\hspace{.5mm}h^{(0) \mu\nu}
(\mathbb{G}^{{\ell}}h^{(0)})_{\mu\nu} \hspace{.5mm}+\cr \nonumber\\
&&\hspace{.8cm}+\mathbb{A}_m\left\{h^{(m) \mu\nu}
(\mathbb{G}^{\ell}h^{(m)})_{\mu\nu}-\frac{1}{4}\mathbb{M}^{2}\bigg( h^{(m)\mu\nu}
h^{(m)}_{\mu\nu}-(h^{(m)})^{2}\bigg)\right\}\bigg],\nonumber\\
\eea
where
\bea\label{rrr}
\mathbb{A}_0&=&
\sigma -\frac{1}{2m^2\ell^2}
-\frac{\kappa}{m\ell} ,\cr
\mathbb{A}_m&=&
\sigma +(\frac{3}{2}-2a)\frac{1}{m^2\ell^2}
-\frac{\kappa\, a^{2}}{m\ell}\cr
\mathbb{M}^2&=&-\frac{1}{\ell^2}\;\frac{(a-1)^2}{m^2\ell^2\mathbb{A}_m}
\eea
One then should impose the following conditions to get a consistent theory
\be\label{unitary}
\mathbb{A}_0>0,\;\;\;\;\;\;\;\mathbb{A}_m>0,\;\;\;\;\;\;\;\mathbb{M}^2\geq-\frac{1}{\ell^2}.
\ee
Using the same procedure as we have used in the body of paper we can find the central
charge of the dual theory. Again in this case  we get
\be
C=\frac{3\ell}{2G} \mathbb{A}_0.
\ee
It is evident from this expression that to get a consistent model $\sigma$ cannot be negative and 
therefore one should set $\sigma=1$. One then should look at the expressions 
\eqref{rrr} to see
 whether there is a range of parameters where both $\mathbb{A}_0$ and $\mathbb{A}_m$ are 
 positive.  Actually using {\it e.g.} Mathematica 
  it is easy to see that in this case the parameters will not 
 satisfy the conditions given in \eqref{unitary} and therefore for $\bar{\sigma}=-1$ one cannot 
 get a consistent model.


\end{document}